\documentclass[fleqn]{2023SCGE}
\usepackage{aas_macros}
\usepackage{amssymb,amsmath}
\usepackage{xspace}
\usepackage{multirow,array,appendix}
\usepackage[percent]{overpic}
\usepackage{xcolor}
\usepackage{graphicx}
\usepackage{tikz}
\usetikzlibrary{arrows.meta}
\usepackage{booktabs} 
\usepackage{siunitx}

\newcommand{\sinc}{{\rm sinc}}
\newcommand{\cube}{\textsc{CUBE}\xspace}
\newcommand{\cubetwo}{\textsc{CUBE2}\xspace}

\UseRawInputEncoding

\begin{document}

\ensubject{subject}

\ArticleType{Article}
\SpecialTopic{}
\Year{2026}
\Month{June}
\Vol{69}
\No{6}
\DOI{10.1007/s11433-025-2926-0}
\ArtNo{269511}
\ReceiveDate{December 3, 2025}
\AcceptDate{January 30, 2026}
\OnlineDate{February 28, 2026}

\title{\cubetwo : A Parallel $N$-Body Simulation Code for Scalability, Accuracy, and Memory Efficiency}{\cubetwo : A Parallel $N$-Body Simulation Code for Scalability, Accuracy, and Memory Efficiency}

\author[1]{Hao-Ran Yu}{{haoran@xmu.edu.cn}}
\author[1]{Bing-Hang Chen}{}
\author[2,3]{Kun Xu}{}
\author[1]{Ming-Jie Sheng}{}
\author[3,4]{Jiaxin Han}{}
\author[5]{Yipeng Jing}{}
\author[6]{Huahua Cui}{}

\address[1]{Department of Astronomy, Xiamen University, Xiamen, Fujian 361005, China}
\address[2]{Center for Particle Cosmology, Department of Physics and Astronomy,
    University of Pennsylvania, Philadelphia, PA 19104, USA}
\address[3]{Department of Astronomy, School of Physics and Astronomy, Shanghai Jiao Tong University, Shanghai, 200240, China}
\address[4]{State Key Laboratory of Dark Matter Physics, Key Laboratory for Particle Astrophysics and Cosmology (MOE), \text{\&} \\Shanghai Key Laboratory \text{\&} for Particle Physics and Cosmology, Shanghai Jiao Tong University, Shanghai 200240, China}
\address[5]{State Key Laboratory of Dark Matter Physics, Tsung-Dao Lee Institute \text{\&} School of Physics and Astronomy, \text{\&} \\Shanghai Jiao Tong University, Shanghai 201210, China}
\address[6]{Dawning Information Industry Co., LTD., Beijing, 100089, China.}

\AuthorMark{Yu H.-R.}

\AuthorCitation{Yu H.-R., et al.}
\abstract{$N$-body simulation serves as a critical method for modeling cosmic evolution and poses a significant challenge in high-performance computing. We present {\cubetwo}, an open-source cosmological $N$-body code emphasizing memory efficiency, computational performance, scalability and precision. The core of its algorithm utilizes  multi-level Particle-Mesh (PM) method to solve the Poisson equation for matter distribution, leveraging the well-optimized Fast Fourier Transform (FFT) for computational efficiency. Precision is ensured by the optimized Green's function that seamlessly bridges gravitational interactions between multi-level PM and Particle-Particle (PP) calculations. The program design enhances per-core/node efficiency in processing $N$-body particles, while the Information Optimized Storage (IOS) addresses memory constraints for large particle counts.  Using {\cubetwo}, we run two cosmological simulations with particle counts of $6144^3$ on the Advanced Computing East China Sub-center (ACECS) to test performance and accuracy.}

\keywords{${\bm N}$-body simulation; cosmology; Large scale structure of Universe; High performance computing}

\PACS{95.35.+d, 98.65.-r, 98.80.-k}

\maketitle
\begin{multicols}{2}

    \section{Introduction}\label{sec.intro}
    The $N$-body problem studies the gravitational dynamics of $N$ particles. When $N \geq 3$, it lacks general analytical solutions and can only be solved numerically using computers, known as the $N$-body simulation.

    Based on the properties of Cold Dark Matter (CDM) and primordial cosmic perturbations, the $N$-body simulation can conveniently model the evolution of the Universe and provide unique insight in the study of large-scale structure.
    For example, by labeling $N$-body particles with unique identifiers and acquiring their complete phase-space information, we are able to trace the evolutionary history of cosmic structures, calibrate redshift space distortion effects, and establish the relationship between cosmic initial conditions and low-redshift observables \cite{massconcentrationredshift_ludlow_2014, 25_reid_2014,modelling_jennings_2010, effect_hellwing_2016, flexible_contreras_2021}.
    By adjusting initial conditions and cosmological parameters, their individual and combined effects on large-scale structure are explored, leading to an optimized cosmic initial condition reconstruction and cosmological parameter interpretation \cite{nonlinear_zhu_2017, isobaric_wang_2017, elucid_wang_2014, differentiable_li_2024}.
    $N$-body simulation is also the starting point in producing mock galaxy catalogs and the basis upon which hydrodynamical simulations of galaxy formation and comprehensive astrophysical processes are developed \cite{ezmocks_chuang_2015, comparing_blot_2019, halogen_avila_2015, semianalytical_tan_2025}.

    The $N$-body simulation is an important application in the field of high-performance computing.
    Interestingly, the release of the fastest supercomputers often accompanies the largest cosmological $N$-body simulations \cite{largescale_angulo_2022, differential_yu_2017}.
    In these simulations, the most time-consuming process is the calculation of gravity between $N$-body particles.
    The simple calculation of the forces between all pairs of particles (referred to as Particle-Particle, PP) has a computational complexity of $O(N^2)$.
    When $N$ becomes large, PP is impractical and approximations are introduced to reduce complexity.
    The Particle-Mesh (PM) method \cite{highperformance_harnois-deraps_2013, accurate_xu_2021, computer_hockney_1988}, taking advantage of the Fast Fourier transform (FFT), and the tree algorithm, using the ``Barnes \& Hut tree'' algorithm \cite{hierarchical_barnes_1986, cosmological_springel_2005, gadget4_springel_2022}, reduce the complexity to $O(N\log N)$.
    Algorithmic combinations can further reduce computational complexity to near $O(N)$, such as the two-level PM scheme employed by \textsc{PMFAST}, \textsc{CUBEP$^3$M}, and \cube , as well as hybrid approaches like \textsc{GADGET-4}, \textsc {PKDGRAV3}, and \textsc{photo$N$s2}, which integrate the fast multipole method with tree-based or PM algorithms \cite{optimal_merz_2005, highperformance_harnois-deraps_2013, cube_yu_2018, gadget4_springel_2022, pkdgrav3_potter_2017, hybrid_wang_2021}.
    To ensure accuracy under these approximations, short-range interactions need to be compensated by PP or by expanding the tree structure to every particle.

    The next-generation galaxy surveys, such as Dark Energy Spectroscopic Instrument (DESI), the Large Synoptic Survey Telescope (LSST) and the China Space Survey Telescope (CSST), cover vast cosmological volumes and resolve faint galaxies, enabling promising studies of key cosmological phenomena including dark energy, primordial non-Gaussianity, neutrino mass, etc. \cite{desi_collaboration_2016, introduction_collaboration_2025, lsst_ivezic_2019, jiutian_han_2025}. These require cosmological simulations with large volume coverage and high mass resolution, which translates to an extremely large problem size quantified by the total number of particles, $N$. This leads to several computational challenges:
    \begin{itemize}
        \item First, given limited computing resources, what is the largest problem size $N$ we can achieve? This is directly related to the memory efficiency.
        \item Second, when total computing resources are increased, typically by adding more compute nodes, can a memory-limited problem scale proportionally in size while maintaining constant total computation time? This property is known as {\it weak scalability}.
        \item Third, with the evolution of multi-core architectures, if a fixed per-node problem size is allocated more processors, will computation time decrease inversely with increased computing power? This is known as {\it strong scalability}.
        \item Finally, it is essential to optimize simulation accuracy without compromising computational efficiency.
    \end{itemize}

    In this paper, we present a new $N$-body simulation code \cubetwo to optimize the above considerations. Its algorithm contains adaptive multi-layer PM and PP method.
    \cubetwo is open source, written in {\textsc Coarray FORTRAN} (CAF). CAF simplifies communication statements between computing nodes to the greatest extent, without using message passing interface (MPI). In addition, two layers of shared memory parallelization are implemented using OpenMP directives.
    The only external library required is FFT, which can be found on almost all computing platforms. \cubetwo uses minimal amount of memory and storage among all $N$-body codes, making it possible to use modest platforms to compete over trillion-particle runs. \cubetwo aims to achieve portability between computing platforms. As an example, \cubetwo used Chinese self-developed supercomputing platform and computing chip to run a series of high-resolution simulations for the CSST cosmological science projects \cite{jiutian_han_2025}. The purpose of this paper is to present an overview and progress in developing cosmological $N$-body applications, and provide a reference to people who might use \cubetwo to run simulations, to do further development and analysis.

    The rest of the paper is structured as follows. \cref{sec.meth} opens with a brief overview of cosmological structure formation paradigm and its $N$-body treatments, followed by three subsections describing methods dealing with force accuracy, parallel scalability, and memory optimization. \cref{sec.results} describes the results from force accuracy tests and CSST simulations. Finally conclusions and discussions are provided in \cref{sec.Conclusion and Discussions}.

    \section{Method}\label{sec.meth}
    According to modern cosmology, primordial perturbations in the early universe evolve under gravity and eventually form the large-scale structure we observe today.
    At low redshifts and on smaller spatial scales, these perturbations become nonlinear and the analytical linear perturbation theory fails to describe their evolution, and we can only rely on numerical simulations.
    At some epochs when the universe has entered the matter-dominated era, and when linear theory is still valid on scales of consideration, we set the initial conditions of the simulation.
    After that, and on subhorizon scales, the total matter content (CDM and baryons) can be modeled by the Newtonian approximation of Einstein-Boltzmann equations.\footnote{A small fraction of cosmic neutrinos remain relativistic components but contribute minimally. They gradually decelerate during cosmic evolution, become non-relativistic (treated equivalently to dark matter). For \cubetwo's neutrino implementation, see Chen et al. Universe 11, 212 (2025) \cite{cosmological_chen_2025}.}
    By discretizing the phase space of matter into $N$-body particles, and a time integrator updating their comoving positions and velocities along transformed time variables, the equations recover Newtonian kinematics and dynamics.
    A comoving volume $V$ satisfying {\it periodic boundary conditions} is commonly used, whose expansion is solved by the Friedmann equations.
    The standard cosmological $N$-body simulation framework neglects small initial relative velocities between CDM and baryons \cite{relative_tseliakhovich_2010} and small-scale baryonic effects at late epochs.

    \begin{figure}[H]
        \includegraphics[width=0.9\linewidth]{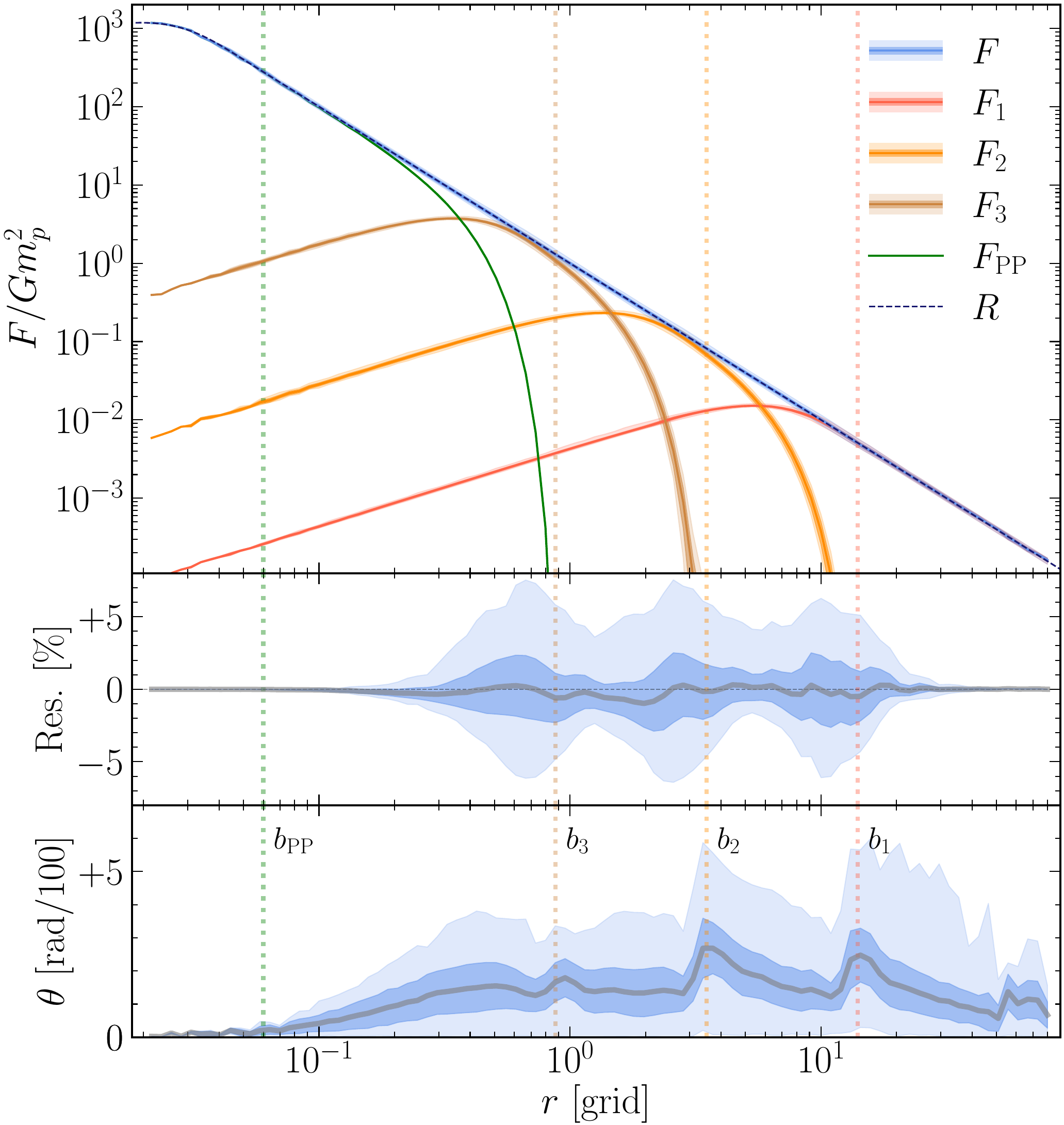}
        \caption{Reference force and its decomposition. Upper panel: the total reference force ($b_{\rm PP}=0.06$\,{[grid]}) is decomposed into four components, each calculated by \cubetwo (discussed in \cref{sec.Force accuracy}), with shaded regions denote force errors. Middle and lower panels: magnitude residual and directional error between computed force sum $\bm F$ and true reference force $\bm R$. In all three panels, inner error regions indicate $1\sigma$ standard deviation, and outer regions indicate the minima to maxima of the distribution.}
        \label{fig.pairwise}
    \end{figure}

    \subsection{Gravity Calculation} \label{sec.Gravity Calculation}
    \subsubsection{Reference force decomposition} \label{sec.Reference force decomposition}
    The force calculation is the most time-consuming part of an $N$-body code.
    Upon Newtonian approximations, the gravitational force between particles follows an inverse-square law.
    However, to avoid nonphysical particle scattering, the force between very close particles needs to be suppressed, a procedure known as {\it softening}. For instance, one may artificially add a small constant $b$ to any interparticle distance, $r \rightarrow r + b$ , though this globally introduces a systematic bias. Alternatively, defining a softening length $b$ where the force vanishes for $r < b$ mimics the gravity of a spherical shell of radius $b$ on test masses. A more generalized approach assigns isotropic density profiles $S(r)$ to all particles; for $r<b$, the overlapping of profiles softens the force, gradually decaying to zero, i.e., $F(r \to 0)=0$.\footnote{We use bold letters for vectors (such as ${\bm F}$), and use regular letters for scalars and the magnitude of vectors ($F\equiv|{\bm F}|$).} Common profiles include uniform spherical shells, solid spheres, or Gaussian distributions. We adopt
    \begin{equation}\label{eq.denprofile}
        S(r,b) =
        \begin{cases}
            48\left(b/2 - r\right)/\pi b^4 & r < b/2     \\
            0                              & r \geq b/2,
        \end{cases}
    \end{equation}
    which gives better accuracies in 3-dimensional (3D) case \cite[Eq.\,(8-2)]{computer_hockney_1988}. The corresponding gravitational reference force\footnote{Since gravity is an attractive, radial force, for simplicity we can just express the magnitude. Also, since in cosmological simulations specific units are used, we neglect the coefficients in the inverse square law.}
    \begin{equation}\label{eq.refforce}
        R(r,b) =
        \begin{cases}
            \frac{64r}{5b^3} - \frac{256r^3}{5b^5} + \frac{32r^4}{b^6}                                      \\ \ \ \ \ \ \ + \frac{1536r^5}{35b^7} - \frac{192r^6}{5b^8}&  0 \leq r < b/2 \\
            \frac{3}{35r^2} - \frac{32}{5b^2} + \frac{256r}{5b^3} - \frac{96r^2}{b^4} + \frac{256r^3}{5b^5} \\ \ \ \ \ \ \ + \frac{32r^4}{b^6}  \quad - \frac{1536r^5}{35b^7} + \frac{64r^6}{5b^8}&  b/2 \leq r < b  \\
            \frac{1}{r^2} & r \geq b.
        \end{cases}
    \end{equation}
    An example of the reference force is shown by the dashed curve in the upper panel of \cref{fig.pairwise}.

    \cubetwo assumes a total reference force $R(r,b_{\rm PP})$,  where the softening $b_{\rm PP}$ is usually a small fraction (many simulations use $0.1$) of the average particle spacing $H_p\equiv(V/N)^{1/3}$. It is further decomposed into a local PP term and $M$ layers of PM terms,
    \begin{equation}\label{eq.force_decomp}
        R_{\rm tot}\equiv R(r,b_{\rm PP})=R_{\rm PP}(r)+\sum_{{\alpha}=1}^M R_{\alpha}(r),
    \end{equation}
    where $R_{\rm PP}(r)$ is computed by the local PP method, and $R_{\alpha}(r)$ represents PM force labeled by ${\alpha}$. The force calculation of each component should closely approximate its reference force, with their sum approaching the total reference force. The decomposition \cref{eq.force_decomp} is arbitrary, and is based on the principle that the force softening is carried out on scales of the PM grid to minimize the error. Most naturally, the PM employs a same softening, \cref{eq.refforce}, with the softening radius $b_{\alpha}$, corresponding to a softening caused by a density profile $S(r,b_{\alpha})$. \cubetwo uses three layers of PM ($M=3$), thus
    \begin{subequations}\label{eq:all_PM}
        \begin{align}
            R_1(r)             & = R(r,b_1)  \label{eq.PM1}                                \\
            R_2(r)             & = R(r,b_2) - R(r,b_1) \label{eq.PM2}                      \\
            R_3(r)             & = R(r,b_3[\varrho]) - R(r,b_2) \label{eq.PM3}             \\
            R_{\mathrm{PP}}(r) & = R(r,b_{\mathrm{PP}}) - R(r,b_3[\varrho]), \label{eq.PP}
        \end{align}
    \end{subequations}
    where $R_1,R_2,$ and $R_3$ are reference forces corresponding to PM method on a global coarse mesh (PM1), a local mesh with fixed resolution (PM2), and a finer mesh with adaptive resolution (PM3). The softening scales satisfy $b_1>b_2>b_3[\varrho]>b_{\rm PP}$. $b_3[\varrho]$ depends on the local clustering, $\varrho$, see \cref{sec.Local force and strong scaling}.
    More over, $R_2,R_3,$ and $R_{\rm PP}$ are {\it truncated} beyond distances $b_1,b_2,$ and $b_3[\varrho]$. Taking $R_2$ as an example, since $R_1(r>b_1,b_1)=R_{\rm tot}$, $R_2(r>b_1) = 0$, i.e., $R_2$ is truncated at $b_1$. These truncations give the virtue of calculating forces locally. The only global force is $R_1$, without truncation.
    The upper panel of \cref{fig.pairwise} shows an example of the above decomposition.
    The dashed curve represents the total reference force ($b_{\rm PP}=0.06$\,{\small{[grid]}}), and the rest curves, calculated by \cubetwo, closely approximate each reference force components and their sum (\cref{sec.Force accuracy}).

    \subsubsection{PM Force Optimization} \label{sec.PM Force Optimization}
    PM treats force and potential as fields that cover the entire simulation space, with the references field being the negative gradient of the potential field. The four fundamental steps of PM are 1) assigning particle masses to a mesh to obtain the density field, 2) solving the Poisson equation to derive the potential field, 3) taking the gradient to obtain the references force field, and 4) interpolating the accelerations from mesh back to particles.
    To ensure that the PM method closely approximates the reference forces Eqs. (\ref{eq.PM1}\,-\,\ref{eq.PM3}), it is necessary to optimize the PM calculation strategy, as well as its kernel functions, i.e., the Green's functions $G_\alpha$.
    We now derive $G_\alpha$ under fixed constraints 1), 3), and 4).

    For mass assignment, we employ Triangular Shape Cloud (TSC) interpolation, i.e., convolving the mass distribution $m(\bm x)\equiv m_p\sum_{i=1}^{N_p}\delta(\bm x - \bm x_i)$ of $N_p$ particles with mass $m_p$ with the 3D TSC distribution function
    \begin{equation}\label{eq.filter}
        W(\bm{x})\equiv \prod_{d=1}^3    \begin{cases}
            3/4-x_d^2               & |x_d|<1/2        \\
            \Big(3/2-|x_d|\Big)^2/2 & 1/2\leq|x_d|<3/2 \\
            0                       & |x_d|\geq 3/2,\end{cases}
    \end{equation}
    then sampled on the mesh to obtain the density field $\rho({\bm n})\equiv\rho({\bm  x})|_{\bm n}=\left[m(\bm x)*W(\bm x)\right]|_{\bm n}$, where for simplicity, we set the length of the grid $H=1$, and $\bm n \equiv (n_1,n_2,n_3)\, $ is the 3D integer vector indicating the grid number of the mesh.
    This parameterization constrains $n_d$ ($d=1,2,3$) within $1 \leq n_d \leq n_g$ by definition, where $n_g$ specifies the number of grid cells per dimension. For its corresponding Fourier wave-vector, $\bm{k} \equiv (k_1,k_2,k_3)$, has a periodicity of $k_g = 2\pi /H = 2 \pi$.
    To solve the Poisson equation for the gravitational potential, we transform to Fourier space, which simplifies the convolution between density field and the Green's function $ \phi({\bm n}) = \rho({\bm n}) * G_\alpha({\bm n})$ to a multiplication of their Fourier counterparts, $\phi(\bm{k}) = \rho(\bm{k})G_\alpha(\bm{k})$.
    The force field ${\bm E}({\bm n})$ on the mesh is approximated by a four-point finite difference of the potential:
    \begin{eqnarray}
        {\bm E}({\bm n}) \equiv E_d({\bm n})=\bm D({\bm n})*\phi({\bm n}),
    \end{eqnarray}
    where $\bm D({\bm n})\equiv \sum_{d=1}^{3}D(n_d)\hat{\bm n}_d$ is the difference operator, with
    \begin{eqnarray}\label{eq.diff}
        D(n)=\frac{4}{3}\frac{\delta(n+1)-\delta(n-1)}{2}-\frac{1}{3}\frac{\delta(n+2)-\delta(n-2)}{4},
    \end{eqnarray}
    providing fourth-order numerical accuracy. Finally, to ensure momentum conservation, we interpolate the force field to particles using the same distribution function, ${\bm F}(\bm x)=W(\bm x)*{\bm E}({\bm n})$. Finally, the force of the particle at ${\bm x}_2$ due to the particle at ${\bm x}_1$ is
    \begin{equation}\label{eq.F_alpha}
        \bm{F}_\alpha=\sum_{\bm k} W{\bm D}G_\alpha\sum_{\bm n} W({\bm k}_{\bm n}) e^{-i {\bm k}_{\bm n} \cdot \bm{x}_1} e^{i{\bm k} \cdot{\bm x}_2},
    \end{equation}
    where the inner summation defines ${\bm k}_{\bm n} = {\bm k}+2\pi{\bm n}$, to correct for the alias effects due to finite resolution. $W$ is the Fourier transform of the TSC distribution \cref{eq.filter},
    \begin{equation}
        W(\bm{k}) =\left[\prod_{d=1}^3\sinc({k_d}/{2})\right]^3,
    \end{equation}
    where $k_d \in [-\pi,\pi]$ are the components of $\bm k$, and $\sinc$ is the sine cardinal function. The finite difference $\bm D({\bm n})$ is written in Fourier space as
    \begin{equation}
        {\bm D}({\bm k}) = i \sum_{d=1}^3\, \left(\,\frac43\sin k_d-\frac16\sin2k_d\, \right),
    \end{equation}
    where $i$ is the imaginary unit.

    We optimize $G_\alpha({\bm k})$ to minimize the force error from PM discretizations.
    PP calculation does not contribute error on all scales. For PM, when the particle separation is much larger than the grid length, the force is also accurate.
    Thus, we can regard the error dominated by PM at short distances as comparable to grid length.
    Following the procedures in \cite{accurate_xu_2021}, minimizing the variance of the PM force error $\int{\rm d}{\bm {x}_1}\int{\rm d}{\bm x} \left|\bm{F}_{\alpha}(\bm{x};\bm{x}_1)-\bm{R}_{\alpha}(\bm{x}) \right|^2$ leads to
    \begin{equation}\label{eq.Gk}
        G_{\alpha}({\bm k}) = \frac{\bm{{D}}(\bm{k})\cdot{\sum_{\bm n}}\,W^2(\bm{k_n}){\bm R}^*_{\alpha}(\bm{k_n})}{|\bm{{D}}(\bm{k})|^2\left[{\sum_{\bm n}}\,W^2(\bm{k_n})\right]^2},
    \end{equation}
    where the superscript $*$ denotes the complex conjugate and the reference forces in Fourier space are written as
    \begin{equation}\label{eq.Rk}
        {\bm R}_{\alpha}(\bm{k}) = -i \bm{k}\frac{S^2(k,b_{\alpha})-S^2(k,b_{{\alpha}-1})}{k^2},
    \end{equation}
    with $S(k,b)$ being the Fourier transform of \cref{eq.denprofile},
    \begin{equation}\label{eq.Sk}
        S(k,b) = \frac{12}{(kb/2)^4} \left( 2 - 2 \cos \frac{kb}{2} - \frac{kb}{2} \sin \frac{kb}{2} \right).
    \end{equation}
    Note that in \cref{eq.Rk}, $b_{\alpha}, b_{{\alpha}-1}$ are the softening and truncation distances of the PM considered. We also have to set $b_0=\infty$, leading to the fact that PM1 does not need to be truncated.

\begin{figure}[H]
        \centering
        \includegraphics[width=0.45\textwidth]{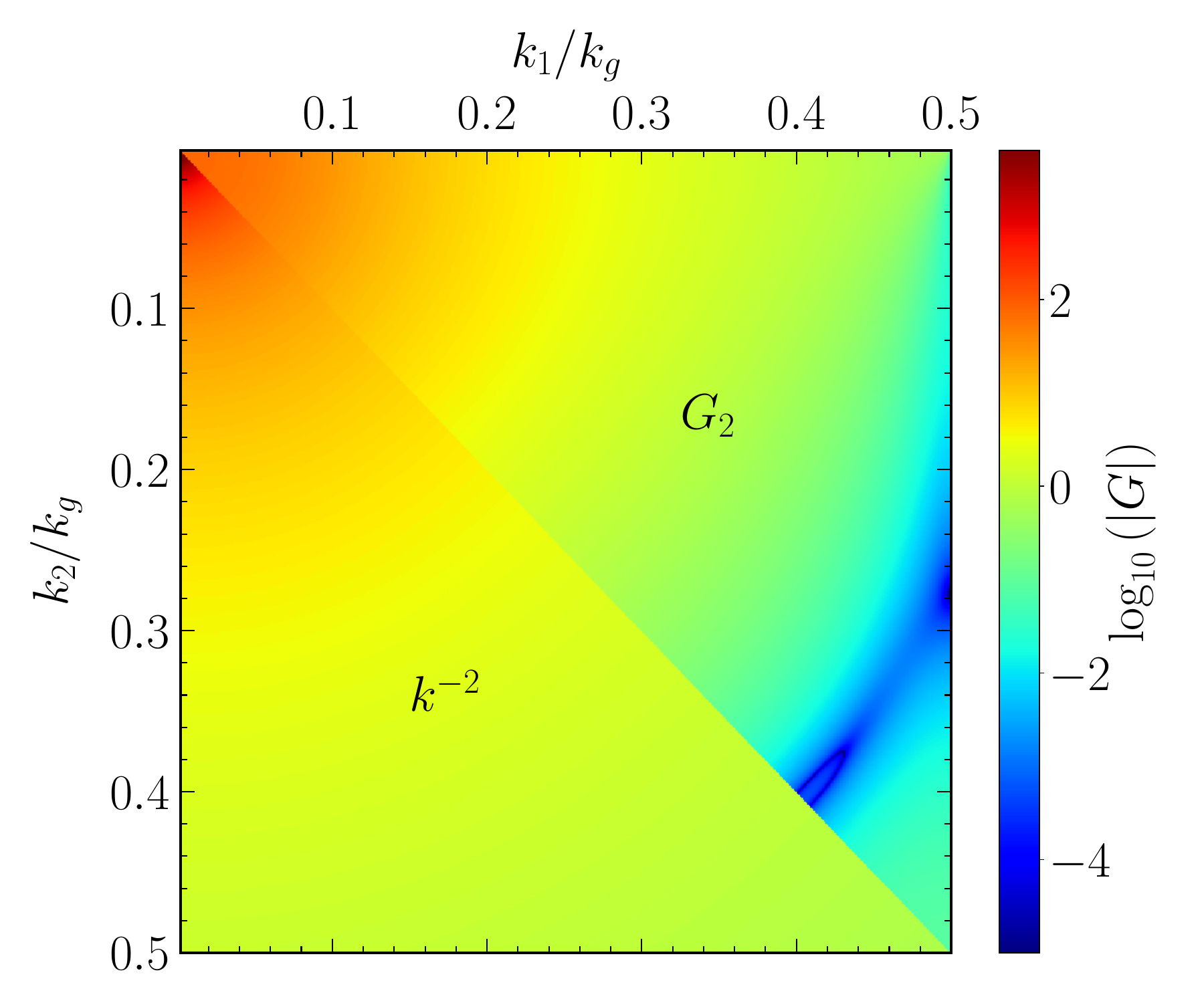}
        \caption{Optimized Green's function $G_2$ (upper triangle, symmetric between $k_1$ and $k_2$) and $k^{-2}$ (lower triangle) for comparison, on the $k_1k_2$-plane with $k_3=0$.}     \label{fig.Gk}
    \end{figure}

    \cref{fig.Gk} plots the Green's function $G_2$ at $k_3=0$ (upper triangle), compared to the suboptimal, continuous case $1/k^2$ (lower triangle). They match only on intermediate scales, between softening $b_2$ and truncation $b_1$. $G_2$ decays at larger scales to match PM1, and on small scales, the features correspond to mass assignment effects, finite differencing, distribution functions and aliasing corrections.

    \cref{eq.Gk} involves infinite summations over $\bm n$. In fact, all variables decay sufficiently rapidly as $|\bm n|$ increases, so in practice we can just sum up to $\max(|\bm n|)\leq 2$. Although the calculations might still be time-consuming, these Green's functions do not need to be computed at each time step, but only once before simulation starts. The Green's functions are the same when the simulation resolution is unchanged, so they can be stored as files for restarting the simulation or different simulations with same resolutions.

    \subsection{Spatial decomposition and scaling} \label{sec.Spatial decomposition and scaling}
    \subsubsection{Spatial decomposition} \label{sec.Spatial decomposition}

    The code is structured to realize the multi-level force calculation. In nonshared-memory architectures, the way particles are stored plays a decisive role in performance and scalability.
    Common methods include irregular volume partition based on fractal space-filling curves, usually Hilbert curves, and regular volume partitions, such as slab, pencil, or cubic decompositions.
    In our case, FFT in the PM algorithm favors regular-shaped volumes. Furthermore, local PM calculations and particle send/receive processes need additional buffer zones surrounding the boundaries of the local subvolumes.
    Among the above partition methods, cube has the smallest surface-area-to-volume ratio, which can minimize additional calculations and memory consumptions.
    Thus \cubetwo adopts a multi-level cubic decomposition of the simulation volume, henceforth its name.

    \begin{figure}[H]
        \centering{\includegraphics[width=0.4\textwidth]{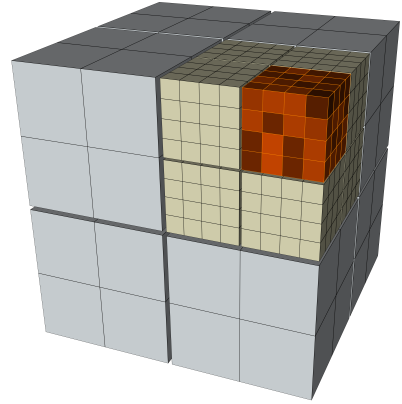}}
        \caption{Illustration of the \cubetwo volume decomposition hierarchy. For each dimension, the box size contains 2 nodes (gray), each node contains 2 tiles (yellow), and each tile contains 4 subtiles (red). The color variations in subtiles indicate different PM3 grid resolutions.}\label{fig.cubes}
    \end{figure}

    \cref{fig.cubes} gives an illustration of volume decomposition of \cubetwo.
    In this example, the global cosmological volume is decomposed into $8$ sub-cubes (except the closest one, the rest $7$ sub-cubes are painted as gray); each of them is usually stored in a computing node, handled by an MPI process. Subsequently, within each node, the sub-cube is decomposed into $8$ tiles (except the closest tile, the rest $7$ tiles of the closest sub-cube are painted as yellow); although they share memory, only one tile at a time is undergone local (PM2) force calculation and position update. Furthermore, each tile is decomposed into $64$ subtiles, and on each subtile we calculate PM3 and PP. The resolution of PM3 (and thus the PP range) is adaptive, depending on the matter clustering in the subtile (the subtiles of the closest tile are painted as red, and darker tones represent more clustering).

    \subsubsection{Global force and weak scaling} \label{sec.Global force and weak scaling}
    The only force to be computed globally is PM1, which is the only part having a complexity of $O(N\log N)$. Cutting down the computational load in this part helps keep the total complexity close to $O(N)$.

    The two free parameters of PM1 to be tuned are the resolution and softening. The grid size of the PM1 mesh, $H_1$, is coarsen to $r_1$ times the average particle spacing: $H_1=r_1 H_p$,
    and PM1's softening scale $b_1$ is set to $\beta_1$ times this grid size: $b_1=\beta_1 H_1$.
    \cubetwo sets $r_1=4$ and $\beta_1=3.5$ by default, balancing accuracy, speed, and memory efficiency.
    Larger $r_1$ reduces the memory consumption of PM1 fields (Green's function, density, potential), global communication, and calculation of global FFT by a factor of $r_1^{-3}$, which is suitable for parallel systems with low communication speed, and/or for extremely large particle number $N$.
    Larger $\beta_1$ improves the force matching accuracy between PM1 and PM2. In practice, $\beta_1\gtrsim 3$ yields reasonable results for accurate simulations, while $\beta_1\gtrsim 4$ provides marginally better force matching, but the simulation results already converge. Since PM1 softening is compensated by PM2 (whose truncation length is $b_1=r_1\beta_1 H_p$), increasing either $r_1$ or $\beta_1$ intensifies the PM2 computation and memory load. The same logic also applies to PM2 and PM3.

    The actual PM1 calculation starts from generating coarse grid density fields on each computing node, using TSC interpolation. Due to the geometry of TSC distribution function, additional one layer of buffer zone surrounding the sub-cubes is used to transfer the density information to adjacent nodes. The global, distributed FFT of the density field should then be applied on the PM1 mesh. There are many distributed FFT libraries that can be incorporated into the code.

    By default, \cubetwo uses the Coarray features to transpose and rearrange FFT arrays from cubic decomposition to {\it pencil} decompositions. In this way, each computing node governs a pencil beam of array which is continuous over one global direction, such that a local 1D FFT can be done on that direction. After rearranging and Fourier transforming the data over three directions, the 3D Fourier transform is completed, and the Fourier harmonics $\rho(\bm{k})$ are stored in a pencil-based format. The advantage of pencil decomposition is that the parallelization can scale to more number of nodes.\footnote{
        For example, if PM1 mesh resolution is $N_1^3$, distributed over $N_n^3$ nodes, in cubic, pencil, slab decompositions, the arrays take the dimension $
            (N_1/N_n,N_1/N_n,N_1/N_n)$, $(N_1,N_1/N_n,N_1/N_n^2)$ and $(N_1,N_1,N_1/N_n^3)$ respectively. In the case of slab decomposition, $N_1/N_n^3$ must be an integer, which is more difficult to satisfy than the pencil case.
    }

    PM1's Green's function $G_1({\bm k})$ is also computed in a pencil-based structure and optionally stored in files. Each node only needs to compute/read the harmonics it governs before the $N$-body main iterations. After multiplying $\rho(\bm{k})$ with $G_1({\bm k})$ on each node, the potential is inverse Fourier transformed to real space, stored as cubic decomposition. Two layers of buffer region are needed for the four-point finite difference computing to obtain the force field. The force is then interpolated to the particles to update their velocities.

    The complexities of the above calculations are all linear except the Fourier transform, which is $O(N_1\log N_1)$, where $N_1$ is the PM1 grid number. It also involves a global communication, arranging data between cubic and pencil decompositions, which might not be linear depending on the topology of the parallel computing system. Nonetheless, these overheads are mitigated by coarsening the PM1 grid, i.e., the PM1 grid number much smaller than the particle number $N_1=r^{-3}_1N \ll N$. The remaining computations are all local and linear, which will be discussed below. As a result, the overall complexity of the computation is $\epsilon O(N\log N)+O(N)\rightarrow O(N)$. This scaling behavior is further corroborated in practical simulations (see \cref{sec.simulations,sec.scalability}).

    \subsubsection{Local force and strong scaling} \label{sec.Local force and strong scaling}
    The other two layers of PM and PP calculations are local. When the problem size $N$ increases, we just increase the number of nodes in the same proportion, and the cubic decomposition inside a node is unchanged. Thus, the complexities of these local force calculations are all linear. As discussed in the next subsection, the memory optimization enables the storage of large number of particles per node, so it is essential to utilize the multi-core parallelization to speed up the heavy computation, meanwhile keeping the memory footprint low.

    PM2 force compensates for the PM1 softening. We set the grid size of the PM2 mesh to $H_2=H_p$ and the softening length $b_2=\beta_2 H_2$ (by default $\beta_2=3.5$). PM2 needs a buffer zone surrounding each tile to guarantee the correctness of the force calculation. The depth of the buffer is sufficient if one takes into account the truncation $b_1$ plus the TSC and finite-difference interpolation range $3H_2$. During the density field construction, particle information from the adjacent tiles (might be from adjacent nodes) should be transferred first, including all possible particles that may exert PM2 forces to the tile. A standard PM algorithm is applied with periodic boundary conditions.
    Although this periodic PM algorithm is applied to a local, non-periodic volume, all the spurious forces induced by the periodic boundary conditions ($r>b_1+3H_2$) are eliminated via PM2 force truncation, $R_2(r>b_1)=0$ (the same logic extends to PM3). After obtaining the force field, only the velocities of physical particles are updated.

    PM3 and PP forces compensate the rest short range part of the total reference force. The PM3 truncation and the PP softening are fixed to be $b_2$, $b_{\rm PP}$. The resolution of PM3 is {\it adaptive} according to the local clustering behavior (denoted by $\varrho$) of the subtile so as to minimize the total computing time of PM3 and PP. For given subtile, let the grid size of PM3 mesh to be $H_3=r_3[\varrho]H_2$. The computation time of PM3, $t_{\rm PM3}$, is dominated by FFT, and a very weak dependency on particle number in the subtile $N_p$, so we assume $t_{\rm PM3}\sim A_1r_3^{-3}\log(r_3^{-3})+\epsilon_1 N_p$.
    Now we consider PP computing time $t_{\rm PP}$. The PP force is truncated at PM3 softening, $b_3=\beta_3 H_3 = \beta_3 r_3[\varrho] H_2\propto r_3$ (by default $\beta_3=3.5$), also called PP range. PP calculation includes constructing the linked list data structure such that the particles are easily indexed. It should cover the subtile and its buffer zone, whose depth is $b_3$. Then we loop over particles inside the subtile and find their particles within the PP range and compute their force according to \cref{eq.PP} and update their velocities. Longer PP range leads to heavier computation. We assume $t_{\rm PP}\sim A_2r_3^3+\epsilon_2 N_p$.

    \begin{figure}[H]
        \includegraphics[width=0.4\textwidth]{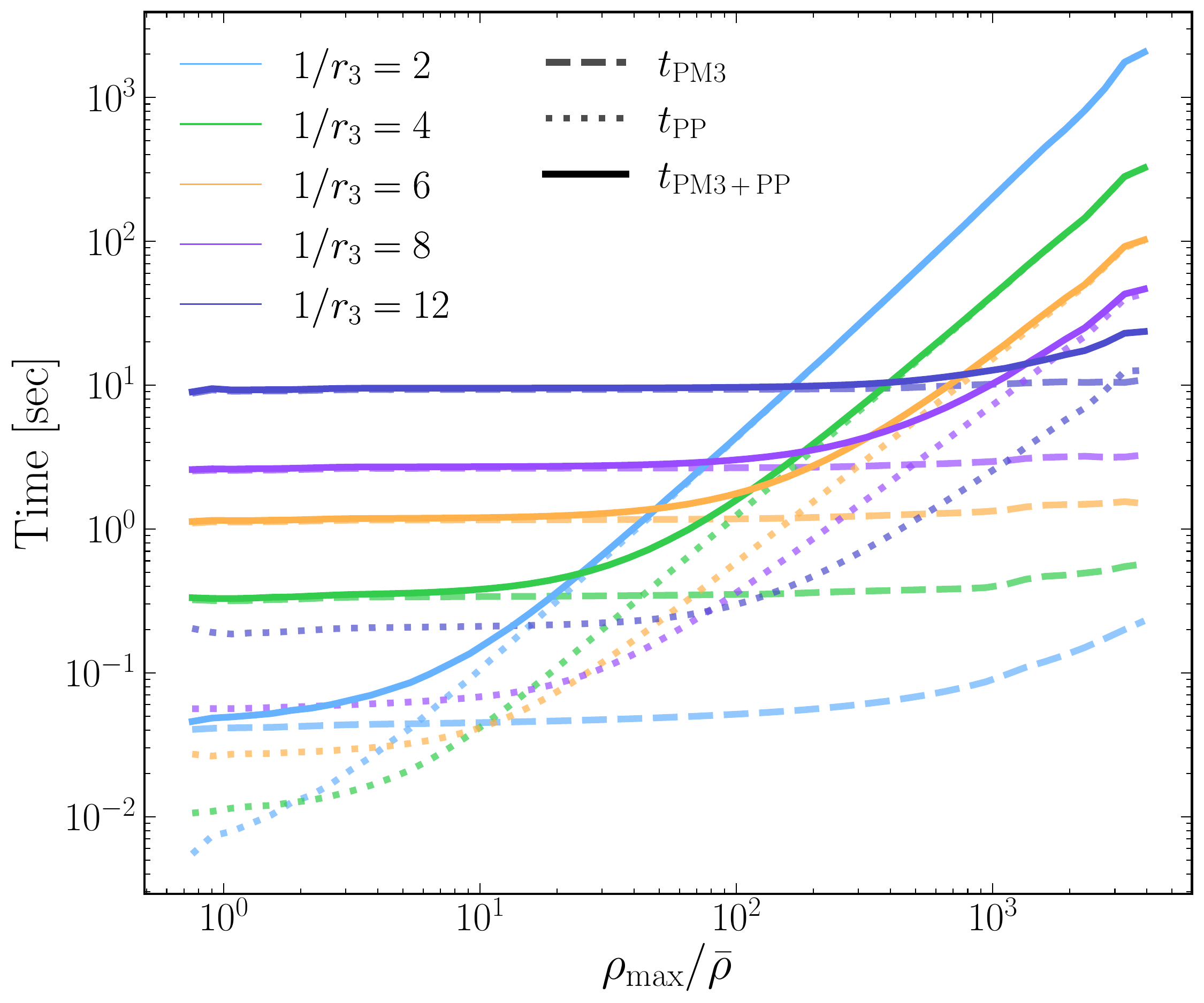}
        \caption{Time consumption comparisons of the adaptive PM3+PP algorithm across refinement levels: PM3 (dashed curves), PP (dotted curves), and total (solid curves) for different resolutions are shown by different colors.}        \label{fig.PMtime}
    \end{figure}

    The subtile typically corresponds to small cosmic scales where density perturbation exhibit large variations. Given $\varrho$, we find $r_3$ to minimize $t_{\rm PM3}+t_{\rm PP}$. The best strategy is to measure the timing in simulations -- \cref{fig.PMtime} presents an example, plotting the average $t_{\rm PM3}$, $t_{\rm PM3}$ and $t_{\rm PM3}+t_{\rm PP}$ as functions of maximum coarse grid density $\rho_{\max}$ of the subtile. Different colors denote PM3 mesh resolutions relative to PM2, i.e., $1/r_3$. Clearly, each $t_{\rm PM3}$ grows slowly with increasing $\rho_{\max}$. This is because at high mesh resolutions, $t_{\rm PM3}$ is dominated by FFT, whereas at low resolutions, processes such as mass assignment contribute discernible fraction of computing time. In contrast, $t_{\rm PP}(\rho_{\max})$ are rapidly increasing functions, but for higher resolutions, $t_{\rm PP}$ exhibits a slower growth rate as $\rho_{\max}$ increases. This means that when $\rho_{\max}$ is large, a high-resolution PM3 should be used, an acceptable additional $t_{\rm PM3}$ trades off the otherwise prohibitive $t_{\rm PP}$. While the optimal mapping $r_3(\rho_{\max})$ could be dynamically optimized at runtime, a pre-calibrated results (e.g., \cref{fig.PMtime}) exhibit near-universal applicability across different mass resolutions and box sizes.

    Although we have minimized $t_{\rm PM3}+t_{\rm PP}$, it is still a steep increasing function of $\rho_{\max}$, leading to a large variation of computing time across subtiles. This {\it load balancing problem} may cause suboptimal {\it strong scaling}. \cubetwo optimizes it by using nested OpenMP parallelization, dividing $N_{\rm core}$ Central Process Unit (CPU) cores into $N_{\rm team}$ teams, and each team contains $N_{\rm mem}=N_{\rm core}/N_{\rm team}$ members of cores.
    When computing PM3 and PP, $N_{\rm team}$ subtiles are simultaneously assigned to $N_{\rm team}$ teams, and each team uses $N_{\rm mem}$ for an inner (nested) parallelization, corresponding to OpenMP loops dealing with PM3 mass/force assignments, FFT, and linked list iterations in PP calculation. The strong scaling of these can be optimal as long as the grid number per dimension is sufficiently larger than $N_{\rm mem}$.

    \begin{figure}[H]
        \centering
        \includegraphics[width=0.48\textwidth]{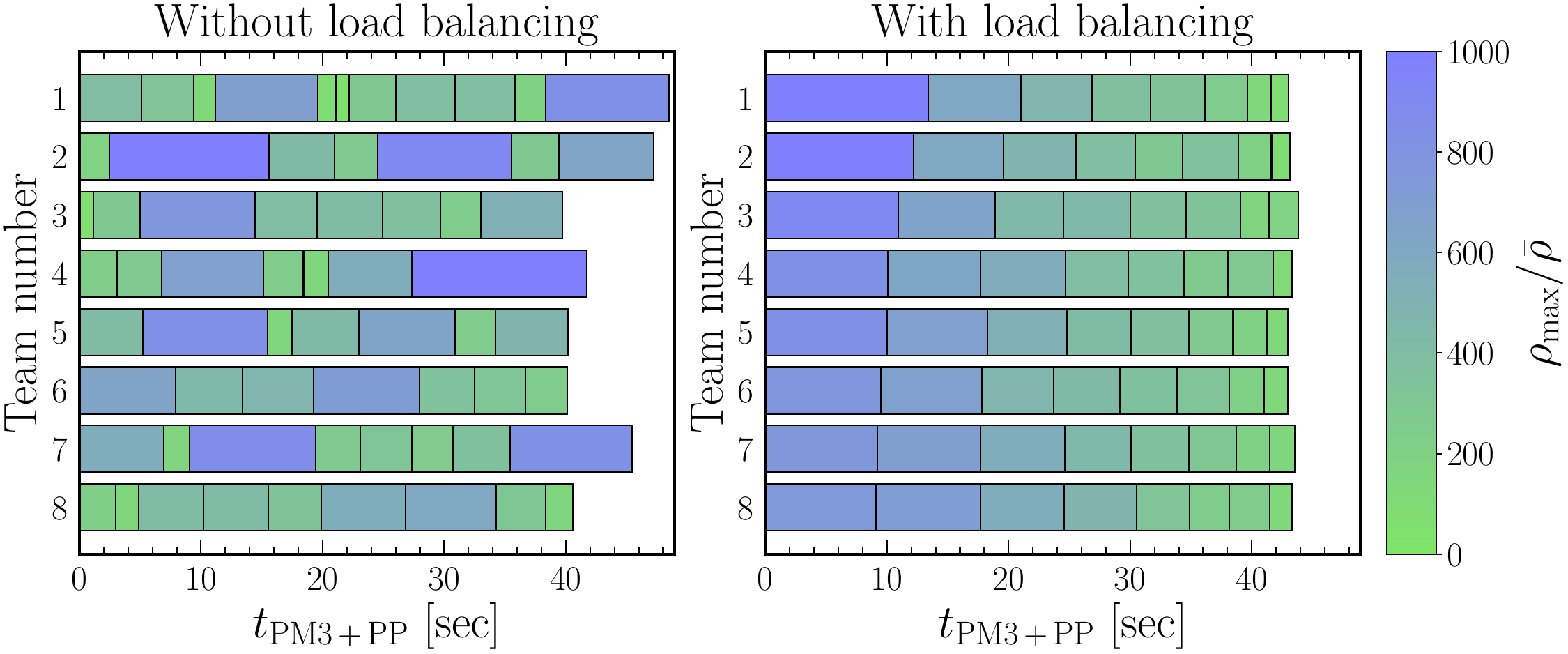}
        \caption{Schematic diagram of load balancing strategy. Left panel: parallel processing in a random order. Right panel: parallel processing after sorting tiles by task size in descending order. We render the colors of tasks according to $\rho_{\rm max}$ -- bluer/greener corresponds to high/low $\rho_{\rm max}$. The optimized approach reduces load imbalance by $11.4\%$ and achieves near-perfect load balancing at $99.6\%$.} \label{fig.balance}
    \end{figure}

    The outer parallelization deals with uneven computation load of subtiles. Sequentially looping over subtiles leads to suboptimal strong scaling. \cubetwo sorts subtiles by $\rho_{\max}$, and subtiles with greater $\rho_{\max}$ are preferentially assigned for an early computation. We show the advantage of this strategy in \cref{fig.balance} -- eight teams deal with the same set of tasks sequentially (left panel) and as the order of decreasing $\rho_{\max}$ (positively but not perfectly correlated to $t_{\rm PM3}+t_{\rm PP}$).
    In the latter case, all teams (threads) finish the computation nearly simultaneously without any thread idling, resulting in a optimized strong scaling.
    We notice that, the strong scaling of the outer parallelization is usually better toward higher $N_{\rm team}$, however it opens up $N_{\rm team}$ copies of temporary memory.
    One can tune the parameters $N_{\rm team}$ and $N_{\rm mem}$ according to the actual scaling tests and memory restrictions.

    Other computations are also easily scaled. PM2 is designed to have similar strategy of PM3, but the computation is conducted on tiles.
    In most simulation configurations, the cosmic scale of a tile is large enough so we do not even need to sort them by $\rho_{\max}$.
    For PM1, it restricts $N_{\rm team}=1$ since there is only one cubic volume to tackle per computing node.
    The most time consuming parts of the computation are global communication (global density field transpose) and FFT.
    The former is easily parallelized by using OpenMP iterations.
    For FFT, since data are stored in a pencil-based format, containing a bundle of 1D global array, and the number of arrays in a bundle is usually much larger than $N_{\rm core}$, many in-place 1D FFTs with same configurations are straightforward to be parallelized.
    Besides updating particle velocities (kick), we also need to update particle positions (drift).
    This part consumes subdominant computing time and is easily parallelized.

    \subsection{Memory minimization}\label{sec.Memory}
    Memory consumption is usually the bottleneck of $N$-body simulations. In particular, we need to store the 6D phase-space coordinates of particles. Traditionally, using 4-byte floating-point numbers, we need at least $24$ bytes per particle (bpp). Based on the Information Optimized Storage (IOS) \cite{cube_yu_2018}, the memory consumption minimizes toward $6$ bpp.

    For positions, a mesh is constructed on the simulation volume, and the particle position relative to the mesh grid is stored by 1-byte-integers (or 2-byte-integers), providing an spatial resolution of $1/256$ (or $1/65536$) of a grid size. An integer-based particle number field is accompanied to infer their global position.
    Similarly, 1-byte-integers (or 2-byte-integers) store the particle velocities relative to the grid, accompanies with a bulk velocity field on the mesh. A formula is needed for the conversion between integers and actual velocities while optimizes the velocity resolution (see \cite[Sec.\,2.2]{cube_yu_2018} for details). To keep the memory footprint low, the time integration adopts a leap-frog method where particle positions and velocities are updated at interlaced times. Higher order time integrators store additional phase-space information and increase the overall memory, but do not significantly increase accuracy.

    IOS can also serve as the linked list data structure, saving another 4 to 8 bpp. The essence of IOS is that the ordering of particles contains a wealth of information but does not require additional memory. For many fast simulations that do not need to resolve subhalo structures, using 1-byte integers is accurate enough. Otherwise, using 2-byte integers gives the full accuracy, but only increases the basic memory consumption to 12 bpp. One can tune the accuracies of positions and velocities separately. The simulation outputs can also use IOS to save disk space.

    \section{Results} \label{sec.results}
    \subsection{Force accuracy} \label{sec.Force accuracy}
    Since PM and PP forces are all additive, we can measure the force error between particle pairs as a benchmark. We create particle pairs randomly onto a grid and use \cref{sec.Gravity Calculation} (with $1/r_3=4$ fixed) to calculate the force components and sum.\footnote{A difference from realistic cosmological simulation is that it replaces the periodic boundary conditions with the isolated boundary condition (see \ref{Appx.IBC}), so that the reference force at large $r$ comparable to box size $L$ remains $r^{-2}$.}
    Their averages and standard deviations are plotted in the upper panel of \cref{fig.pairwise} as a function of the distance $r$ of the particle pair.
    The total force ${\bm F}={\bm F}_1+{\bm F}_2+{\bm F}_3+{\bm F}_{\rm PP}$ accurately restore the reference force $R$, with their magnitude residual $F/R-1$ and directional error $\theta\equiv\cos^{-1}\left(\bm{F} \cdot \bm{R}/FR\right)$ shown by the middle and lower panels.
    The vertical lines show the force softening $b_{\rm PP},b_3,b_2$ and $b_1$, and as expected, the force errors occur at the latter three, force matching, scales.

    \subsection{Simulations} \label{sec.simulations}
    On Advanced Computing East China Sub-center (ACECS), we scale \cubetwo up to $512$ computing nodes (16384 cores, 32 cores per node), $N=6144^3$ particles, in box sizes $L=2400\,{\rm Mpc}\,h^{-1}$ and $L=1200\,{\rm Mpc}\,h^{-1}$. We denote these two simulations ``S6144-2400'' and ``S6144-1200''. They assume a flat $\Lambda$CDM cosmology, with CDM, baryon, and dark energy density parameter $\Omega_{\rm c}=0.23$,  $\Omega_{\rm b}=0.05$, $\Omega_{\Lambda}=0.72$, the dimensionless Hubble parameter $h=0.70$, as well as $\sigma_8=0.82$, $n_s=0.97$ representing the scalar perturbation amplitude and spectral index. To achieve high accuracy, the reference force softening $b_{\rm PP}$ is set to $0.06\,H_p$.

    For both S6144-2400 and S6144-1200, we use 512 nodes to divide the cosmic volume $V$ into $8^3=512$ sub-cubes, resulting in $768^3$ particles per node in Lagrangian space. Each node is assigned to a MPI-process (although CAF does not use MPI directives). Inside each node, the sub-cube is subsequently divided into $4^3=64$ tiles for PM2 calculation. All 32 cores work on one tile at one time to keep the memory footprint low. Each tile is then divided into $4^3=64$ subtiles for PM3 and PP calculation, and we use $N_{\rm team}=4$ teams, each with $N_{\rm mem}=8$ members for a nested parallelization. The above configuration balances speed and memory consumption on ACECS.

    The initial conditions are set at redshift $z_i=200$, where the power spectrum is given by CAMB \cite{camb_lewis_2011}. The particle positions and velocities follow Zel'dovich approximation \cite{gravitational_zeldovich_1970a} (1st order Lagrangian perturbation theory, 1LPT). Although 2nd and higher order LPTs allow to set $z_i$ lower, they compute and store more fields and use more memory. We choose to use 1LPT at a higher redshift and it was confirmed to give enough accuracy.

\begin{figure}[H]\begin{center}
            \includegraphics[width=0.4\textwidth]{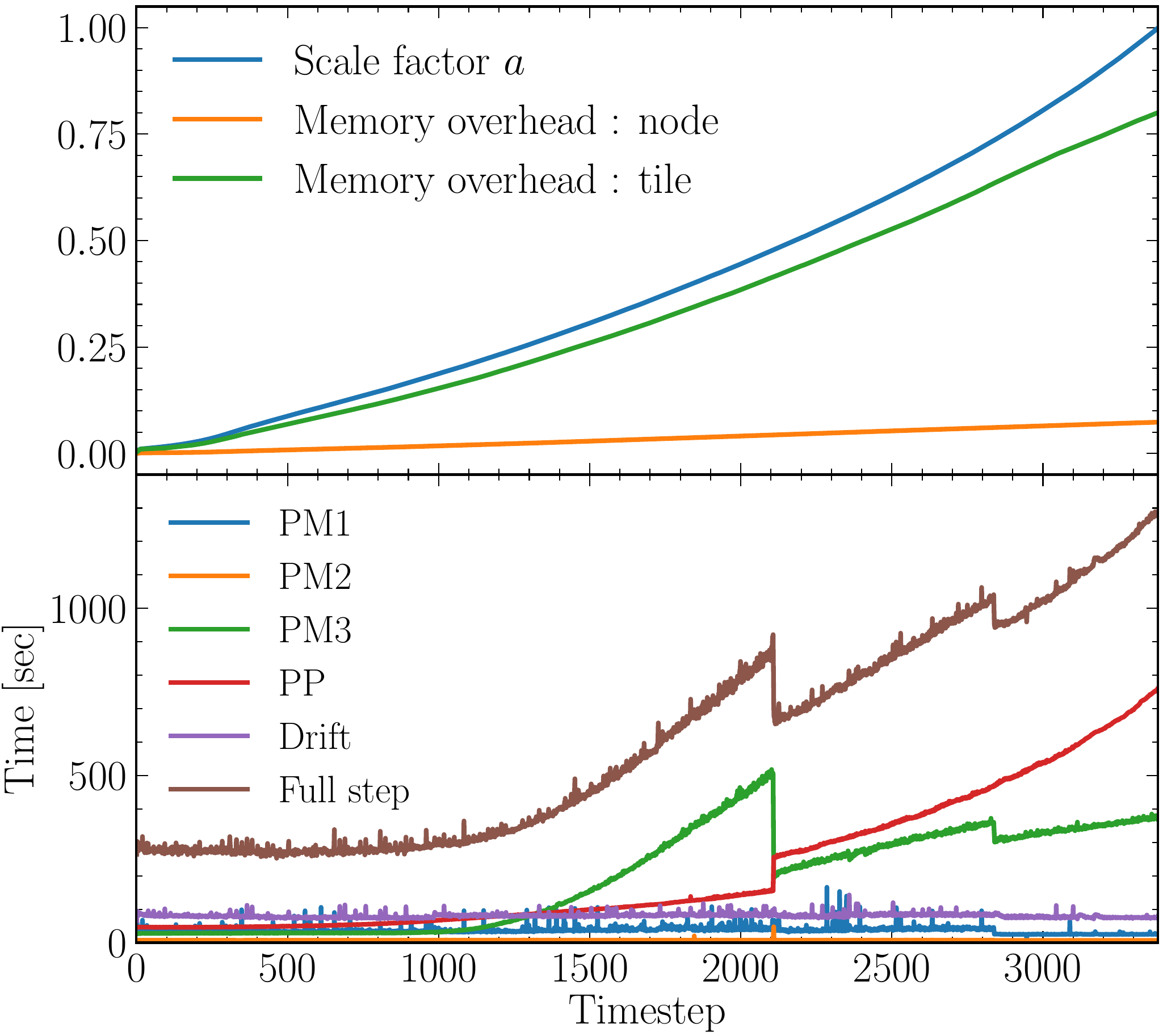}
        \end{center}
        \caption{Performance metrics of  S6144-2400 simulation. Upper panel: growing of cosmic scale factor $a$ as a function of simulation timesteps, as well as the memory overhead on computing nodes and tiles. Lower panel: computing time in different processes at each timestep. Note that we optimized the strategies regarding PP ranges and PM3 resolutions around timestep 2100 and 2800, which reduced the computing time per a full step.}\label{fig.time_2400}
    \end{figure}

\begin{figure}[H]
        \begin{center}
            \includegraphics[width=0.4\textwidth]{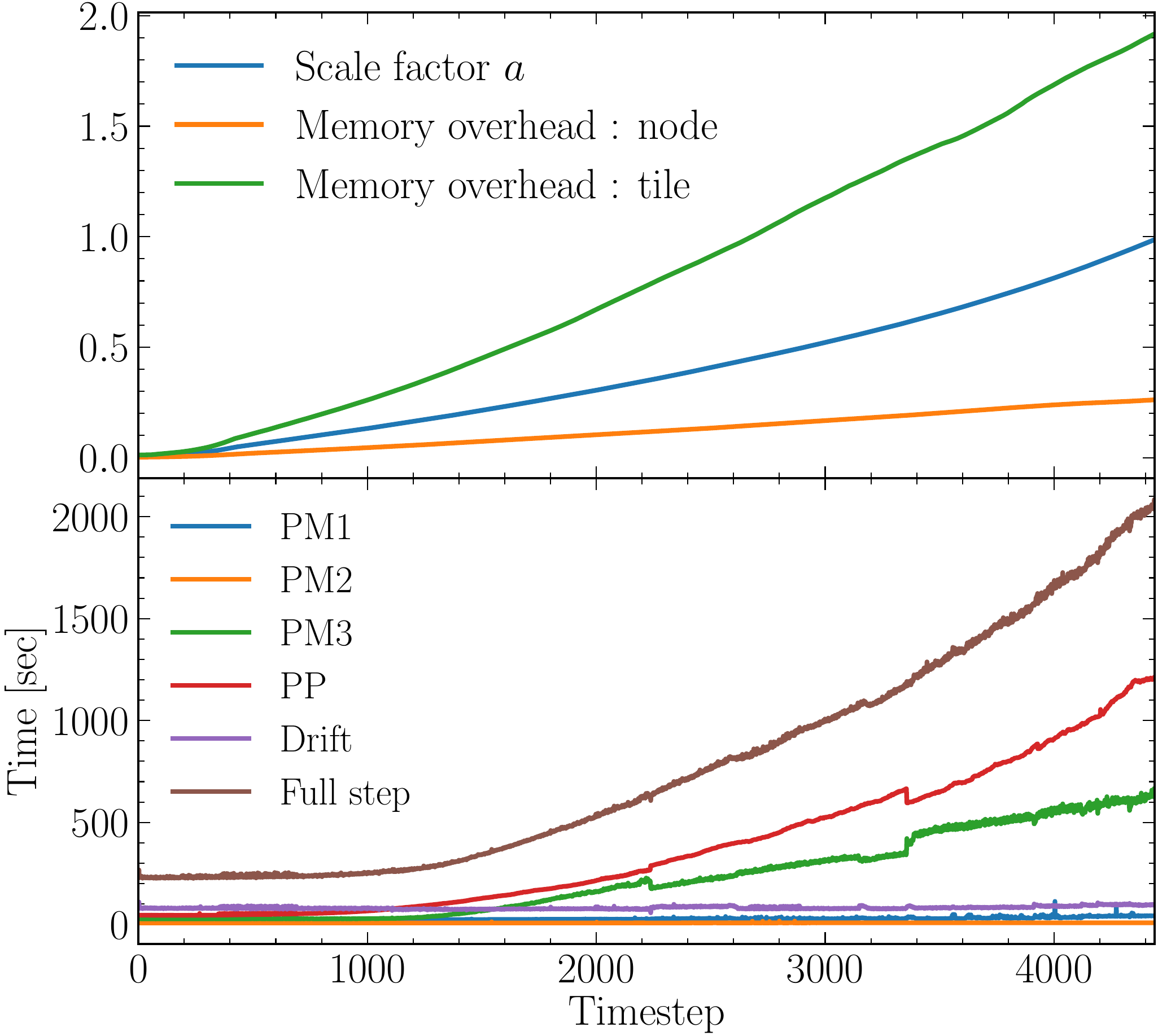}
        \end{center}
        \caption{Same as \cref{fig.time_2400}, but for  S6144-1200 simulation. Note again that we tuned the strategies of PP and PM3 in this simulation around timestep 2100 and 3400, but they did not affect the full step timing significantly.}\label{fig.time_1200}
    \end{figure}

    During the main loop of time integration, the amount of time increment between each timestep $\Delta t$ is constrained by the maximum acceleration and velocity of all particles. The number of total timesteps $N_{\rm step}$ is larger for higher mass resolution (lower $m_p$) and shorter softening $b_{\rm PP}$. In the upper panels of \cref{fig.time_2400,fig.time_1200}, the two blue curves show the grow of scale factor $a$ as a function of timestep for S6144-2400 and S6144-1200.
    These and many other high-resolution simulations find that $\Delta t$ can be empirically set as constant $\Delta a$, and $N_{\rm step}$ is several thousand \cite{cosmicgrowth_jing_2019}.
    In the upper panels of \cref{fig.time_2400,fig.time_1200}, the orange/green curves show the maximum extra portion of memory consumed recorded by nodes/tiles. They arise from the inhomogeneities of the cosmic structure, which is higher for smaller box sizes, and smaller scales (tiles).
    The memory overhead per node limits the maximum number of $N$-body particles can be run. In S6144-2400 and S6144-1200, this memory overhead is controlled below 10\% and 30\%.
    Despite of the higher inhomogeneities on tiles, the actual total memory consumption can be saved in many ways, e.g., reducing the tile sizes, adjusting the nested parallelization, freeing the memory of other computations, etc., while maintaining a high computation performance.


    The lower panels of \cref{fig.time_1200,fig.time_2400} show the time consumption as a function of timestep. As expected, the total time per timestep increases toward low redshift, and is dominated by the force calculation. For both simulations, PM1 consume negligible portion of time, which confirms the complexity of the problem being $O(N)$. PM3 and PP are the most time-consuming parts.    In S6144-2400, around 2100th and 2800th timestep, optimizations to the PM3 strategies were made and reduced the total time. Similar adjustments were made in S6144-1200. The total computation times for S6144-2400 and S6144-1200 are $23.8$ and $40.1$ days. Each of these simulations contains $100$ snapshots evenly distributed in $\log(a)$ space.

    For the final snapshot at redshift $z=0$, we examine the matter power spectrum and the halo mass functions (HMF).

    \begin{figure}[H]
        \centering\includegraphics[width=0.45\textwidth]{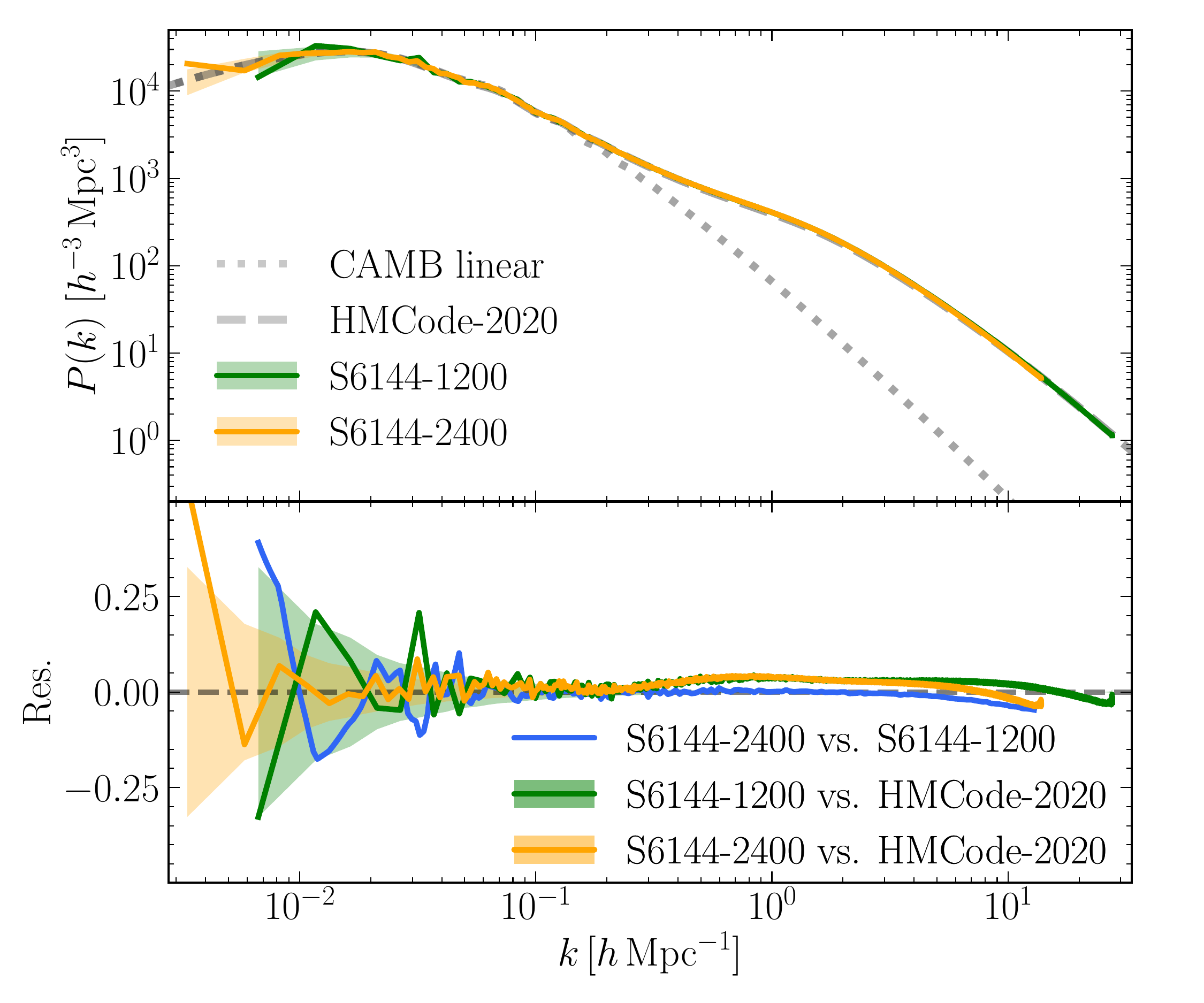}
        \caption{Matter power spectra (upper panel) at redshift $z=0$ for S6144-2400 (orange) and S6144-1200 (green), compared with linear (dotted curve) and nonlinear (dashed curve) predictions. The shaded regions surrounding the nonlinear prediction indicate the cosmic variance arising from the limited cosmic volumes of the two simulations. Lower panel: Residuals of S6144-2400 vs. S6144-1200, and theirs vs. the nonlinear prediction.}\label{fig.power}
    \end{figure}

    We interpolate particles onto a $6144^3$ mesh to obtain the matter overdensity fields, compute the power spectra, and correct for the Poisson noise and aliasing effects using the method of Jing (2005) \cite{jingCorrectingAliasEffect2005}. In the upper panel of \cref{fig.power}, these power spectra are compared with the CAMB linear theory and the nonlinear predictions \cite{camb_lewis_2011,HMCode-2020}. All of them are in good agreement. The lower panel shows the slight residuals. Compared with S6144-1200, the power spectrum of S6144-2400 on very small scales, $k\gtrsim 5\,h\,{\rm Mpc}^{-1}$, is lower by $O(1\%)$ due to the lower mass and force resolutions. On nonlinear scales, $k\gtrsim 0.25\,h\,{\rm Mpc}^{-1}$, the power spectra of both simulations are higher than the HMCode-2020 prediction by $O(5\%)$, but this offset is comparable to the variance between different nonlinear models.
    \begin{figure}[H]
        \centering\includegraphics[width=0.45\textwidth]{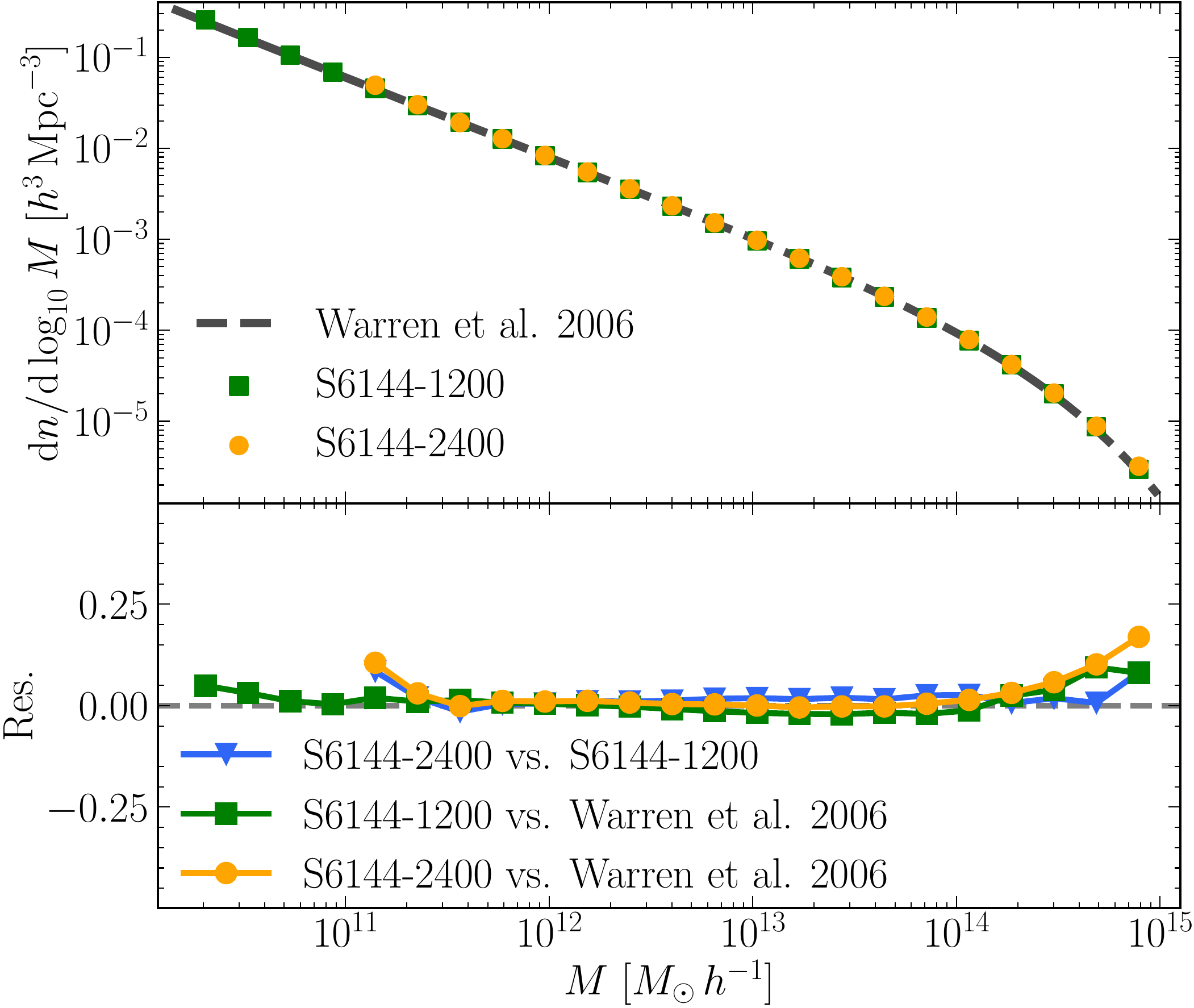}
        \caption{Halo mass functions (upper panel) at redshift $z=0$ for S6144-2400 (orange dots) and S6144-1200 (green squares), compared with the fitting function from Warren et al.\cite{Warren2006} (dashed curve). Lower panel: Residuals of S6144-2400 vs. S6144-1200, and theirs vs. the fitting function.}\label{fig.HMF}
    \end{figure}

    We identify dark matter halos using the standard Friends-of-Friends (FoF) algorithm with linking length set to $0.2H_p$, and we consider all halos with at least $20$ $N$-body particles. In the upper panel of \cref{fig.HMF}, we plot the HMFs of S6144-2400, S6144-1200, and the best-fit model provided by Warren et al. (2006) \cite{Warren2006}. They are in good agreement.
    The lower panel shows the trace residuals. In a wide halo mass range across orders of magnitude, their relative difference is only by $O(1\%)$. As expected, in the mass regions where the numbers of particles in a halo is extremely low, the unbound particles slightly raise the HMF.

    \subsection{Scalability} \label{sec.scalability}
    To test the scalability, additional simulations are conducted.
    First, we fix the problem size per core, downscaling S6144-1200 to 64, 8 and 1 node. Particle numbers and simulation volumes are accordingly scaled. Following our naming convention, they are denoted S3072-600, S1536-300 and S768-150. Compared to the single node S768-150 simulation, the portion of additional time used per timestep (overtime) and weak scaling are shown in \cref{tab:weak_transposed}. Remarkably,  the 512-node problem uses only 6.8\% more time per timestep than a single-node problem, reaching a weak scaling of 94\%.

    Next, we fix the problem size as S768-150, and vary the number of cores $N_{\rm core}$ from 1 to 32. Compared to the single-core test, the speedup (inversely proportional to time elapsed per timestep) due to multi-cores and the corresponding strong scaling are shown in \cref{tab:strong_transposed}. By using 32 cores, the simulation runs 28.1 times faster than a single-core one, reaching a strong scaling of 88\%.
    \begin{table}[H]
        \centering
        \footnotesize
        \setlength{\tabcolsep}{8pt}
        \caption{Weak scaling efficiency of \cubetwo tested on ACECS. The first row shows the simulation name, followed by the number of computing nodes and cores. ``Overtime'' shows the extra amount of time consumed compared to the single-node case. The last row shows the weak scaling efficiency.}\vspace{0.2cm}
        \begin{tabular}{c *{4}{S[table-format=5.8]}} \toprule
            {Simulation}     & {S768-150} & {S768-300} & {S768-600} & {S768-1200} \\ \midrule
            {$N_{\rm node}$} & {1}        & {8}        & {64}       & {512}       \\
            {$N_{\rm core}$} & {32}       & {256}      & {2048}     & {16384}     \\ 
            {Overtime}       & {-}        & {2.6\%}    & {5.1\%}    & {6.8\%}     \\
            {Scaling}        & {-}        & {97\%}     & {95\%}     & {94\%}      \\ \bottomrule
        \end{tabular}
        \label{tab:weak_transposed}
    \end{table}

    \begin{table}[H]
        \centering
        \footnotesize
        \setlength{\tabcolsep}{5.5pt}
        \caption{Strong scaling performance of \cubetwo tested on ACECS. $N_{\rm core}$ shows the number of CPU cores to conduct the simulation S768-150. The last two rows show the speedup ratio compared to the single-core case, and the strong scaling performance.}\vspace{0.2cm}
        \begin{tabular}{ccccccccc}\toprule
            {Simulation} & {} & {}    & {} &   & \hspace{-1.5cm}{S768-150}    & {}   & {}   & {}   \\ \midrule
            {$N_{\rm core}$} & {1} & {2}    & {4}    & {6}    & {8}    & {12}   & {24}   & {32}   \\ 
            {Speedup}        & {-} & {2.0}  & {4.0}  & {5.9}  & {7.8}  & {14.8} & {21.5} & {28.1} \\
            {Scaling}        & {-} & {99\%} & {99\%} & {98\%} & {97\%} & {92\%} & {89\%} & {88\%} \\ \bottomrule
        \end{tabular}
        \label{tab:strong_transposed}
    \end{table}

    \section{Conclusion and Discussions} \label{sec.Conclusion and Discussions}

    We introduce \cubetwo, a parallel $N$-body simulation code designed to achieve optimal weak scalability (\cref{sec.Global force and weak scaling,sec.scalability}), strong scalability (\cref{sec.Local force and strong scaling,sec.scalability}), high force accuracy (\cref{sec.Gravity Calculation,sec.Force accuracy}), and a minimal memory footprint (\cref{sec.Memory}).
    Collectively, these optimizations render \cubetwo highly flexible for diverse $N$-body simulation scenarios. For fast simulations, we can set $b_{\rm PP}$ larger or even disable PP to save computation time, and increase the number of particles per computing node \cite{cube_cheng_2020}. On the other way, for simulations requiring extremely high force resolution, short reference force softening $b_{\rm PP}$ is recommended, paired with increased $\beta_1,\beta_2,$ and $\beta_3$ values to enhance force matching accuracy, and more memory is used trading for faster computing speed. In either case, the high strong scalability ensures the full utilization of all available cores per computing to speed up the computation. Notably, the consistency of the power spectrum and halo mass function (HMF) with established models and across simulations robustly validates the high accuracy of \cubetwo.

    Compared with its predecessor \cube \cite{cube_yu_2018}, \cubetwo has achieved substantial improvements. We summarize the key differences between them as follows.\begin{itemize}
        \item \textbf{PM levels}: \cube uses two-level PM algorithm with fixed resolution ratio. \cubetwo introduces another adaptive PM3 layer, and the resolutions of all the three PM layers are fully adjustable.
        \item \textbf{PM kernels}: \cube uses predefined force kernels \cite{highperformance_harnois-deraps_2013} for its fixed two-level PM. \cubetwo computes optimal Green's function for PM kernels to improve force-matching accuracy.
        \item \textbf{PP force}: \cube lacks a default PP implementation but can integrate PP algorithms (e.g., \cite{highperformance_harnois-deraps_2013}). \cubetwo incorporates a self-consistent PP scheme that is adaptively matched to the PM3 resolution.
        \item \textbf{Force accuracy}: For the maximum relative force error across all possible particle pairs, \cube can exceed $300\%$ (cf. \cite[Fig.\,7]{highperformance_harnois-deraps_2013}). \cubetwo reduces this error by two orders of magnitude, to $7\%$ (root-mean-square $2\%$); see \cref{fig.pairwise}. This improvement stems from higher-order interpolation, finite-difference algorithms, optimal Green's functions, and consistent PP forces. This significantly enhances key statistical results such as power spectra and halo mass functions.
        \item \textbf{Strong scaling}: \cube focuses only on weak scaling and simulations with large boxes. \cubetwo incorporates nested OpenMP parallelization and load balancing strategies, achieving substantial improvements in strong scaling.
        \item \textbf{Miscellaneous}: Communications for global FFT operations are optimized to enhance weak scaling performance. A new parallel FoF halo finder is integrated. The power spectrum estimator incorporates the correction method proposed in \cite{jingCorrectingAliasEffect2005}. Many other bugs are fixed. Notably, \cubetwo retains a concise and readable codebase, with the core code totaling fewer than 3000 lines.
    \end{itemize}

    $N$-body method serves as a fundamental building block for various cosmological simulations. \cubetwo is also equipped with neutrino modules, with massive cosmological neutrinos modeled as particles \cite{TianNu,cosmological_emberson_2017}, multiple fluids \cite{Hydrodynamics} or a background field \cite{cosmological_chen_2025}. Fluid modules, initial condition reconstruction modules, and cosmic angular momentum analysis modules are under development.

    On the optimization front, when the total number of particles $N$ is pushed beyond $10^{12}$ \cite{TianNu,cube_cheng_2020}, the PM1 FFT and global communication load cannot be neglected. In this case, the global PM should be further coarsened, and an additional layer of PM might be required. The localized, time-consuming PM3 and PP calculations are computationally regular and parallelizable, with significant potential for offloading to heterogeneous architectures—a capability that will be developed in the near future.

    \Acknowledgements
    {We thank the anonymous reviewers for their valuable comments and suggestions, which have significantly improved the article. This work is supported by the National Natural Science Foundation of China (NSFC) grant No. 124B2054, 12173030, 12133006, the Fundamental Research Funds for the Central Universities No. 20720240149, China Manned Space Project (No.\ CMS-CSST-2021-A03), National Key R\&D Program of China (2023YFA1607800, 2023YFA1607801), and the Center for international cooperation and disciplinary innovation ("111 Center", Grant No. B20019). The calculation was supported by Advanced Computing East China Sub-center. The code is publicly available on GitHub at \href{https://github.com/yuhaoran/CUBE2}{\tt yuhaoran/CUBE2}.}

    \InterestConflict{The authors declare that they have no conflict of interest.}

    \bibliographystyle{scpma}
    \bibliography{ref}{}

    \appendix
    \section{Isolated boundary conditions} \label{Appx.IBC}
    Under isolated boundary conditions, the reference force $R$ follows the inverse-square law on large scales. Since FFT inherently imposes periodic boundary conditions, we employ a doubled box size ($2L$) \cite{computer_hockney_1988} with modified Green's function to emulate isolation. For a target 3D isolation box length $L$, the reference force is truncated at $r = L$ through
    \begin{equation}\label{eq.iso_appx}
        R_{\mathrm{iso}}(r,b) = \begin{cases}  R(r,b) & r < L     \\  0      & r \geq L, \end{cases}
    \end{equation}
    whose Fourier transform  are written as
    \begin{equation}\label{eq.isok_appx}
        R_{\mathrm{iso}}(\bm{k},b)=\left[1-\mathrm{sinc}(kL)\right]\times R(\bm{k},b).
    \end{equation}

\end{multicols}
\end{document}